# A Possible Complication of Using the 4.14 μm band to Assess D/H Ratios on Icy Bodies

Mark J. Loeffler[1,2], Rebecca A. Carmack[1], Stephen C. Tegler[1],

[1]Department of Astronomy and Planetary Science, Northern Arizona University, Flagstaff, AZ, 86011

[2]Center for Materials Interfaces in Research and Applications, Northern Arizona University, Flagstaff, AZ, 86011

Accepted Astrophysical Journal Letters

Abstract

Determining isotopic ratios on the surfaces of extraterrestrial objects can provide insight into their formation conditions and evolution. Recently, remote-sensing detections of the 4.14 μm HDO absorption band have enabled astronomers to estimate the D/H ratio on several icy objects in the outer solar system. Here, we quantify how ice phase and radiation processing affect the 4.14 μm absorption band. While the band is visible in $H_2O$-ice samples with a crystalline component, we cannot detect it in amorphous $H_2O$-ice. Furthermore, we find that radiation processing, using 10 keV electrons as a proxy, quickly makes this absorption feature undetectable at a rate consistent with the amorphization of an initially crystalline sample. Interestingly, we also find that recrystallizing the irradiated sample causes this band to reappear nearly to its original band depth. For the Saturnian satellite Mimas, we estimate that the HDO feature will decrease by a factor of two within ~$1 \times 10^4$ yrs at typical depths probed by remote sensing spectroscopy. While we suspect that existing methods could still determine the D/H ratio from a surface composed entirely of crystalline $H_2O$-ice, deriving it for surfaces with a significant amorphous fraction may lead to severe underestimation of the true D/H ratio. However, given the direct correlation between the amorphous fraction of the sample and the HDO band depth, we propose that one could still estimate the D/H ratio of a surface with mixed phases by determining the crystalline fraction of the surface $H_2O$-ice using other absorption features.

1. Introduction

Many satellites in the outer solar system show evidence of condensed volatiles on their surfaces (Cruikshank 1976; Cruikshank et al. 1993; Calvin et al. 1996; Noll et al. 1996; McCord et al. 1998; Carlson et al. 1999a; Carlson et al. 1999b; Brown & Calvin 2000; Hibbitts et al. 2000; Grundy et al. 2003; Hansen & McCord 2004; Brown et al. 2006; Grundy et al. 2006; Hansen & McCord 2008; Grundy et al. 2016; Cartwright et al. 2020). On many of these objects, water-ice is estimated to be an important, if not the major, surface component (Calvin, et al. 1996; Carlson, et al. 1999a; Brown & Calvin 2000; Hansen & McCord 2004; Brown, et al. 2006; Grundy, et al. 2006). While $^{1}$H and $^{16}$O are the dominant atoms in water, there are also small amounts of naturally occurring isotopes (e.g. D, $^{17}$O, $^{18}$O). The resulting isotopic compounds, such as HDO, can be detected using infrared spectroscopy, providing a potential avenue for estimating the D/H ratio remotely. Determining a body's D/H ratio can give insight into its formation conditions and evolution (Lecluse et al. 1996; Drake 2005; Yang et al. 2013; Hallis 2017).

Recent observations have confirmed the presence of deuterium by detecting the O-D stretching vibration in HDO at 4.14 μm (Clark et al. 2019; Hedman et al. 2024; Brown et al. 2025; Brown et al. 2026; Tegler et al. 2026). To our knowledge, Clark et al. (2019) made the first attempt to derive the D/H ratio from remote sensing spectra using Cassini VIMS data of Saturnian satellites and rings. Specifically, they estimated the D/H ratio by comparing the band depths of the $H_2O$ absorption at 2.0 μm and the HDO absorption at 4.14 μm against an empirical relation derived from laboratory spectra of $D_2O$ dissolved in $H_2O$ at different concentrations. Brown et al. (2025) applied this approach to JWST observations of seven Saturnian satellites, three of which had not been observed by Clark et al. (2019), and provided leading and trailing hemisphere measurements for five of them. Brown et al. (2026) then used a new technique that relied solely on the HDO band area to assess the D/H ratio on five Uranian satellites. Most recently, Tegler et al. (2026) estimated the D/H ratio on the Uranian moon Titania by comparing JWST data with laboratory-measured optical constants for HDO diluted in $H_2O$.

While this small absorption feature may prove essential to unraveling an icy body's formation conditions, analysis of the 4.14 μm absorption band may be complicated by environmental alterations. For instance, Tegler et al. (2026) recently showed that the strength of this feature in crystalline $H_2O$-ice increases with decreasing temperature. Here, we use a combination of new laboratory experiments and our publicly available laboratory data on the amorphization of crystalline $H_2O$-ice induced by keV electrons (Loeffler et al. 2020) to quantify how ice phase and radiation processing affect the 4.14 μm absorption band. Specifically, we examine whether the band depends on $H_2O$-ice phase and whether radiation processing can alter it on timescales relevant to astrophysical environments. Unless otherwise specified below, we have described all aspects of the laboratory setup, techniques and analysis in detail in our previous work (Loeffler, et al. 2020). We note that the HPLC-grade $H_2O$ used in those experiments contains naturally occurring deuterium, typically ~150 ppm relative to hydrogen.

## 2. Results and Discussion

### *2.1 Infrared spectra of fresh and irradiated $H_2O$-ices*

Figure 1 (top) shows the HDO absorption band (2414 $cm^{-1}$; 4.14 μm) during 10 keV electron irradiation of crystalline $H_2O$-ice at 50 K, compared with that of an amorphous sample (Loeffler, et al. 2020). Figure 1 (bottom) shows the near-infrared spectral region from the same experiments, highlighting the combination and overtone bands of $H_2O$-ice. First, we note that although the 4.14 μm band is clearly present in the crystalline sample (top spectrum in the top panel), it is absent in the amorphous sample (bottom spectrum in top panel). Similarly, absorption features that are associated with crystalline $H_2O$-ice in the near-infrared (top spectrum in the bottom panel), i.e., the peaks at 1.57 μm (6365 $cm^{-1}$) and 1.65 μm (6037 $cm^{-1}$) and a shoulder at 2.06 μm (4835 $cm^{-1}$), are either absent (1.57 and 2.06 μm) or significantly weaker and broader in the amorphous sample (1.65 μm). Irradiation of the crystalline sample causes the area and intensity of 4.14 μm band (as well as the 1.57 and 2.06 μm bands) to decrease until it falls below the noise level, while also causing the 1.65 μm to decrease in height and broaden, until both spectral regions resemble the amorphous sample.

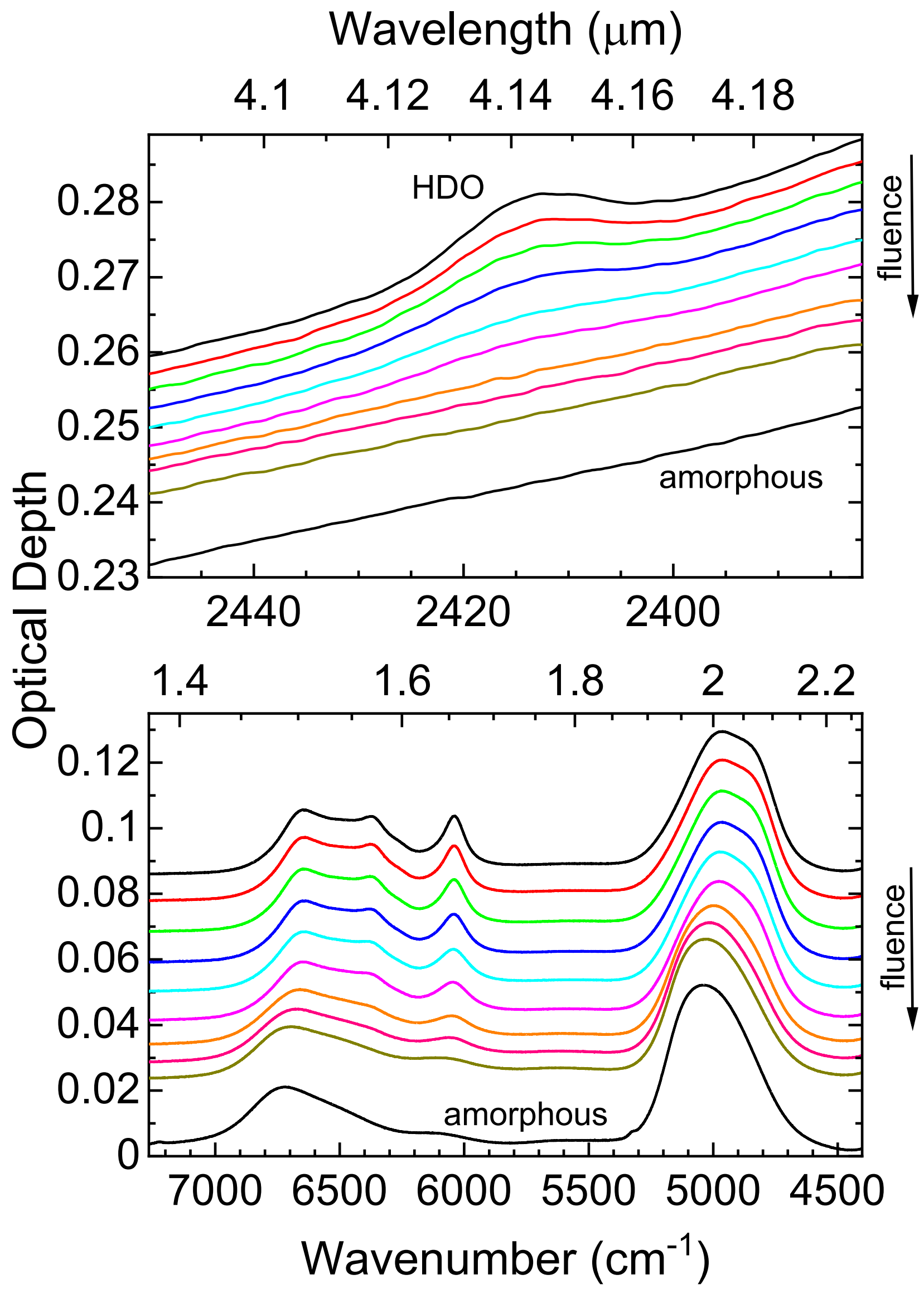

Figure 1. Infrared spectra of the HDO absorption band (top) and $H_2O$ absorption bands (bottom) of a 6.18 x $10^{18}$ $H_2O$ $cm^{-2}$ (2.17 μm) crystalline sample during irradiation with 10 keV electrons at 50 K, compared to an amorphous $H_2O$-ice sample of similar thickness grown at 50 K. The curves (from top to bottom in both panels, displaced vertically for clarity) correspond to fluences of 0, 0.95, 2.61, 6.95, 28.0, 101, 341, 510, and 1590 x $10^{14}$ electrons $cm^{-2}$. The bottom panel has been adapted from Loeffler et al. 2020.

*2.2 The 2.0 and 4.14 μm absorption band depth vs. dose*

The data above qualitatively indicate that the phase of $H_2O$-ice could hinder accurate determination of D/H, whether by using optical constants (Tegler, et al. 2026), directly assessing the HDO band area (Brown, et al. 2026), or by comparing band depth ratios of the 2.0 and 4.14 μm bands (Clark, et al. 2019). To assess this more quantitatively, in Figure 2 we plot the 2.0 and 4.14 μm band depths ($D_B$) from Figure 1 as a function of radiation dose. To determine the band depth, we fit a non-linear continuum to the reflectance spectrum around each absorption feature and applied the following formula (Clark & Roush 1984):

$$D_B = \frac{R_C - R_B}{R_C} \quad (1)$$

where $R_C$ and $R_B$ are the reflectance of the continuum and absorption at the band center, respectively. Figure 2 shows that the band depth of the 2 μm absorption band is relatively insensitive to electron irradiation, whereas the depth of the 4.14 μm band decreases by a factor of two after a dose of ~1 eV $H_2O^{-1}$ and drops down to the noise level by about ~40 eV $H_2O^{-1}$.

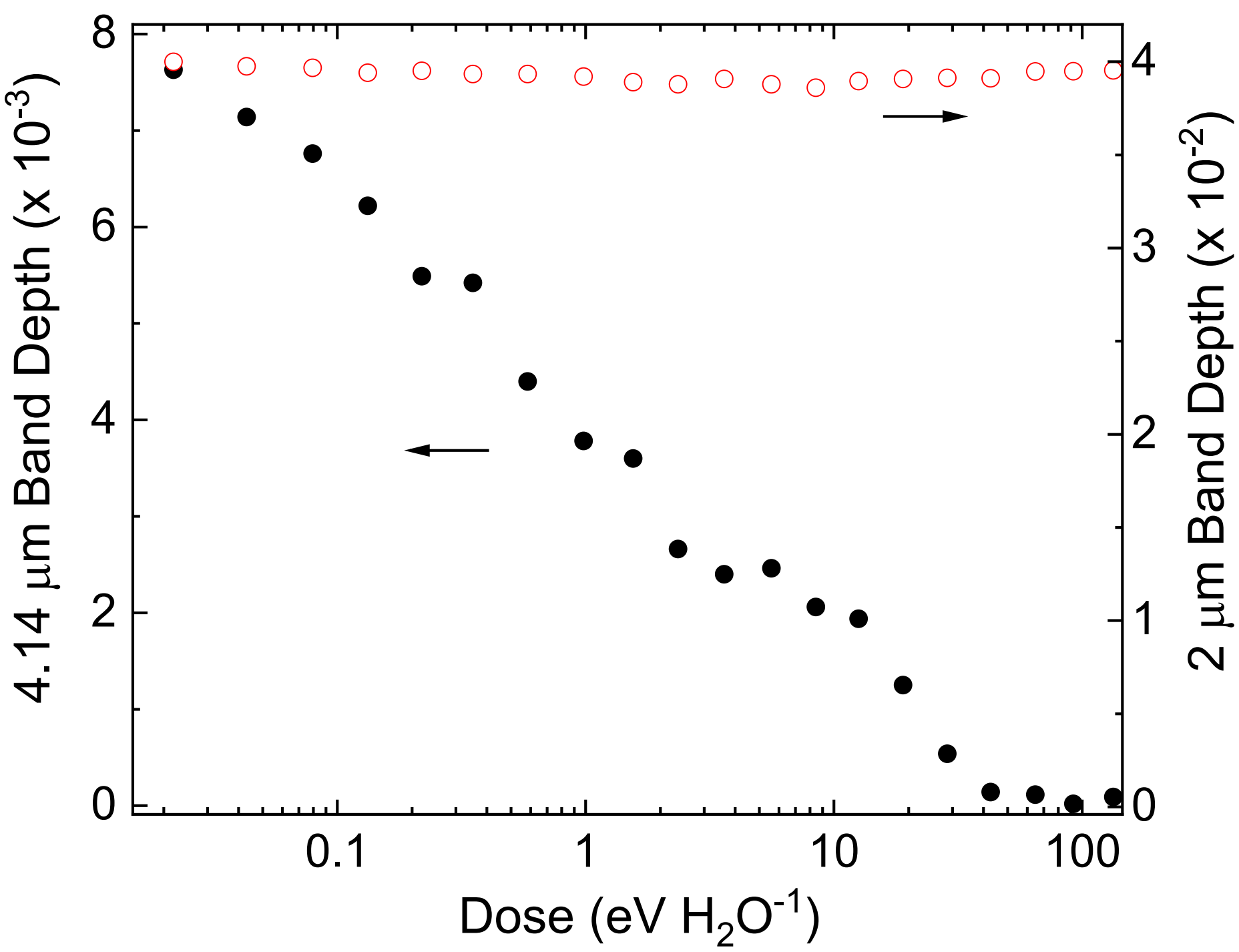

Figure 2. Band depth of the 4.14 μm absorption band (closed circles, left axis) and the 2 μm $H_2O$ band (open circles, right axis) during irradiation of a 6.18 x $10^{18}$ crystalline $H_2O$ cm$^{-2}$ sample with 10 keV electrons at 50 K.

*2.3 Estimating the amorphous fraction of $H_2O$ with the 4.14 μm absorption band*

Loeffler, et al. (2020) shows that the peak to peak of the derivative of the 1.65 and 3.1 μm absorptions can be linked to the amorphous fraction ($\varphi_a$) of a laboratory ice during irradiation following

$$\varphi_a = 1 - \frac{I_p(F) - I_p(a)}{I_p(0) - I_p(a)} \quad (2)$$

where $I_p$ denotes the peak-to-peak intensity of the derivative of a given absorption feature in the spectrum of interest: the initially crystalline spectrum after irradiation at measured electron fluence $(F)$, an amorphous reference spectrum $(a)$, or the unirradiated crystalline spectrum $(0)$. In Figure 3, we show $\varphi_a$ calculated here for the first time using the 4.14 µm feature during irradiation compared to $\varphi_a$ calculated in Loeffler, et al. (2020) for the same sample using the 1.65 µm feature and for a thinner sample using the 3.1 µm feature. The amorphous fraction calculated using all three absorption bands shows a similar trend with radiation dose. Therefore, if the amorphous fraction of an observed surface ice can be determined with the 1.65 or 3.1 µm bands, or another method (Dalle Ore et al. 2015; Dalle Ore et al. 2021), we suspect that the amorphous fraction could be used to scale the measured OD band depth to what it would be on a fully crystalline surface $(D_B(c))$, allowing one to estimate the unaltered D/H ratio. In Figure 4, we plot the Band Depth correction factor ($D_{B,correction}$), or the normalized band depth of the 4.14 µm band, vs. the derived amorphous fraction $\varphi_a$ (Eq. 2). Here, we've expanded the data Loeffler, et al. (2020) collected at 50 K to 10 and 30 K. We did not investigate this relation at higher temperatures, because although electrons can completely amorphize crystalline $H_2O$-ice samples at temperatures as high as 70 K ((Loeffler, et al. 2020), the 10 keV electrons utilized here to process our thick ices are relatively inefficient in damaging the ice. Regardless, at all three ice temperatures studied, the correction factor follows the linear relation, which can be approximated as:

$$D_{B,correction} = 1.0 - \varphi_a \quad (3)$$

The lack of a temperature dependence on the correction factor suggests that one could simply divide the measured OD band depth on a given surface by $D_{B,correction}$ to obtain $D_B(c)$, regardless of the surface temperature.

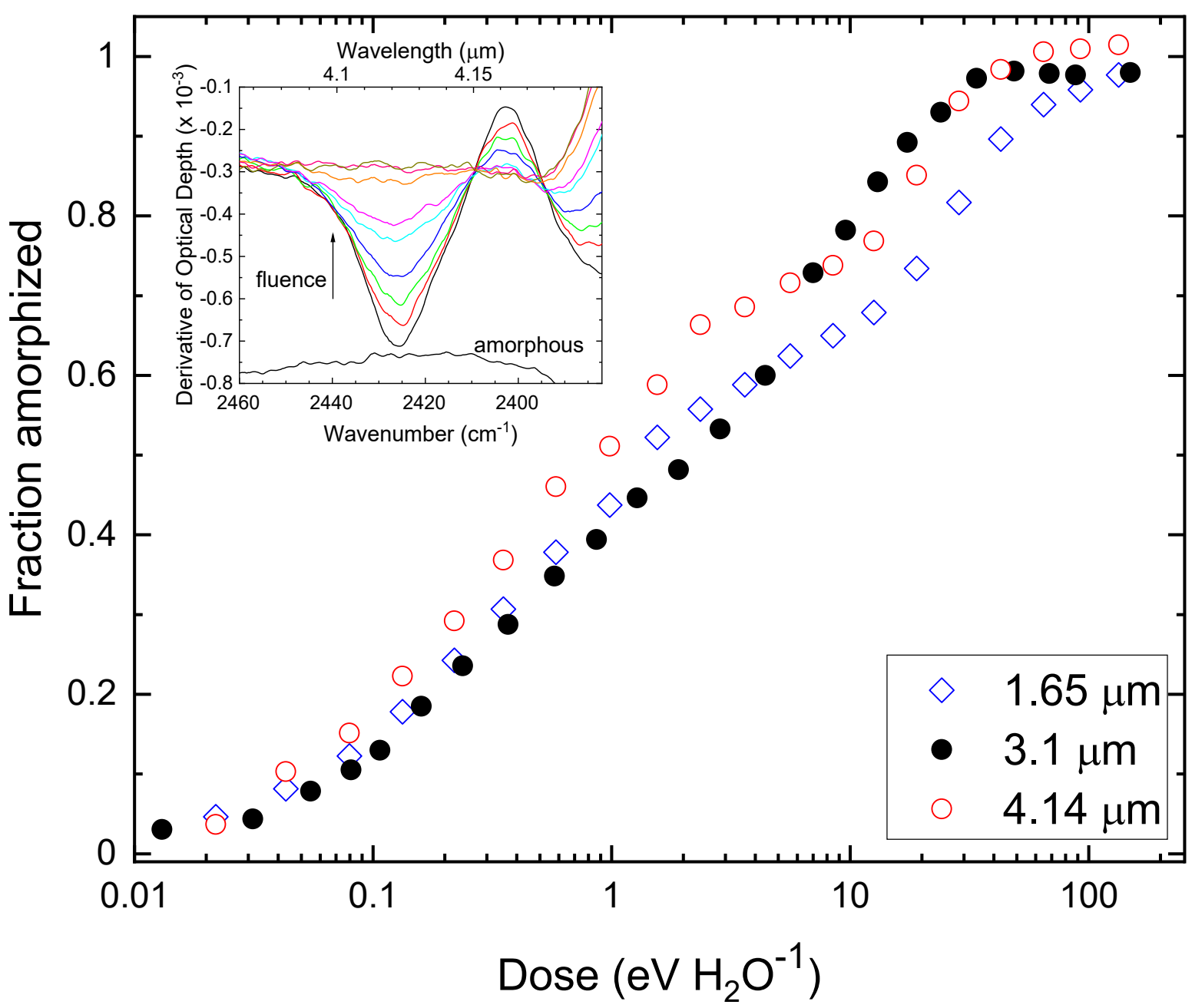

Figure 3. Fraction of amorphous $H_2O$ vs. absorbed dose during irradiation with 10 keV electrons at 50 K determined from the peak-to-peak height of the derivative spectra (see Loeffler et al. 2020 for more details). For the 1.65 (open diamonds) and 4.14 µm (open circles) bands, the sample had a column density of 6.18 x $10^{18}$ $H_2O$ cm$^{-2}$, and for the 3 µm (closed circles) band, the sample had a column density of 1.57 x $10^{17}$ $H_2O$ cm$^{-2}$. Inset: corresponding derivative spectra in the HDO region compared to the spectrum of an amorphous sample; we used these spectra to estimate the amorphous fraction of the sample.

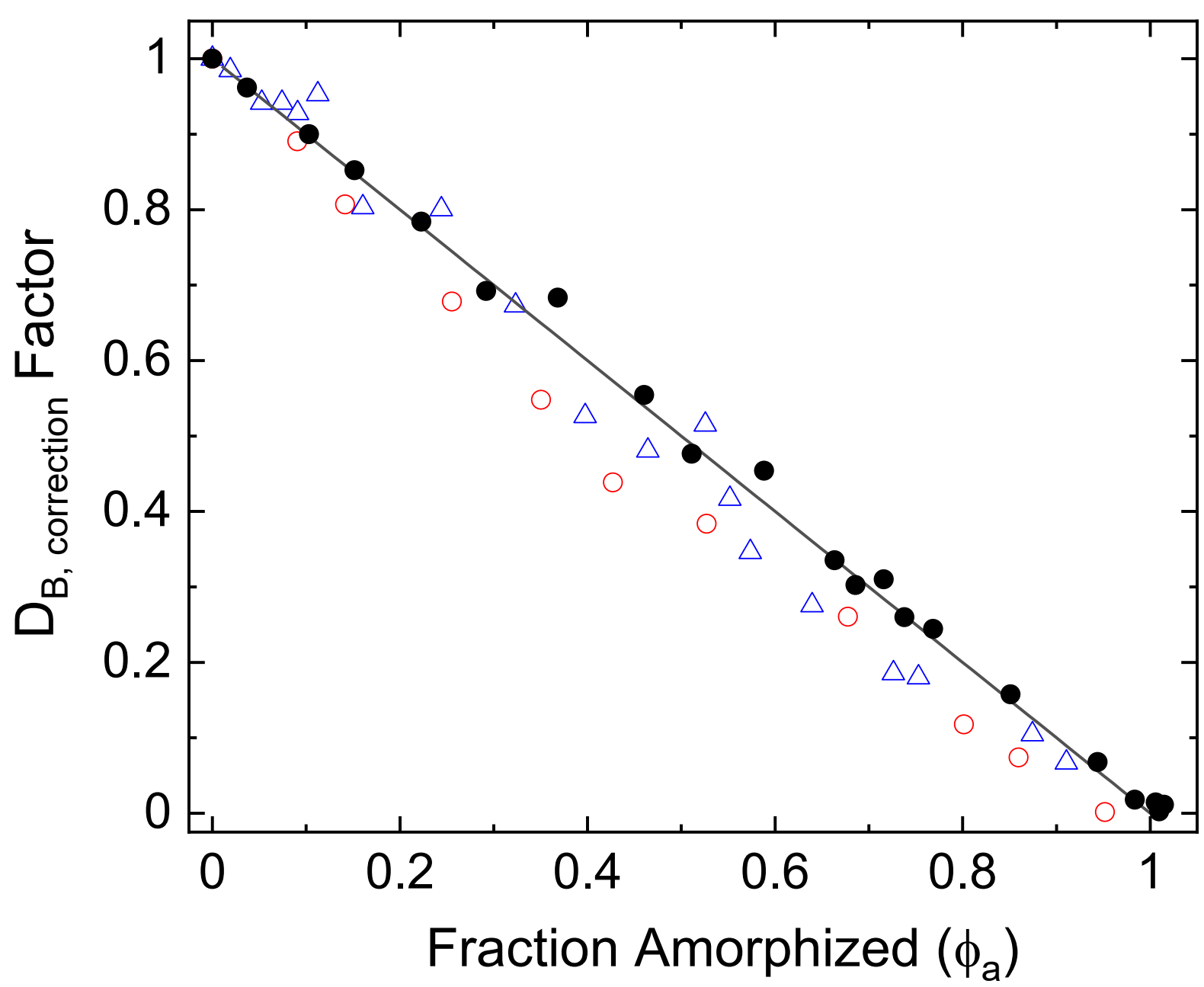

Figure 4. Band depth correction factor, derived from the normalized band depth of the 4.14 μm absorption from Eq. 1, vs. fraction amorphized from Eq. 2 during irradiation of a 6.18 x $10^{18}$ $H_2O$ cm$^{-2}$ sample with 10 keV electrons at 10 K (red open circles), 30 K (blue open triangles), and 50 K (black closed circles). The line is the relation: $D_{B,correction} = 1.0 - \varphi_a$.

*2.4 Recrystallization and the Regeneration of the 4.14 μm absorption*

The data above shows that the 4.14 μm absorption is present in crystalline $H_2O$-ice and decreases during irradiation until it is below the noise level. Here we briefly investigate whether this absorption feature could be recovered if we allow an irradiated sample to recrystallize. Figure 5 shows sample spectra before and after irradiation at 30 K and after heating the irradiated sample to 160 K and recooling it to 30 K. Upon warming, the sample recrystallizes as supported by the reappearance of the 1.65 μm feature (Figure 5 bottom). After recrystallization, the 4.14 μm band is clearly evident but is ~30% weaker compared to the freshly deposited sample and blue shifted by ~0.004 μm (2.5 cm$^{-1}$). We suspect that much of the decrease in intensity is due to a combination of sublimation during heating (we lost ~10% of the sample mass) and breaking of chemical bonds from irradiation, leading to preferential removal of hydrogen (as well as deuterium) and oxygen via sputtering (Bar-Nun et al. 1985; Carmack & Loeffler 2024) and the formation of new species, such as hydrogen peroxide (Moore & Hudson 2000; Loeffler et al. 2006; Zheng et al. 2006; Hand & Carlson 2011). Regardless, the reappearance of the 4.14 μm absorption during crystallization supports that the observed decrease in the band during irradiation has more to do with a change in the crystalline structure of $H_2O$ than in the preferential removal of HDO from the sample.

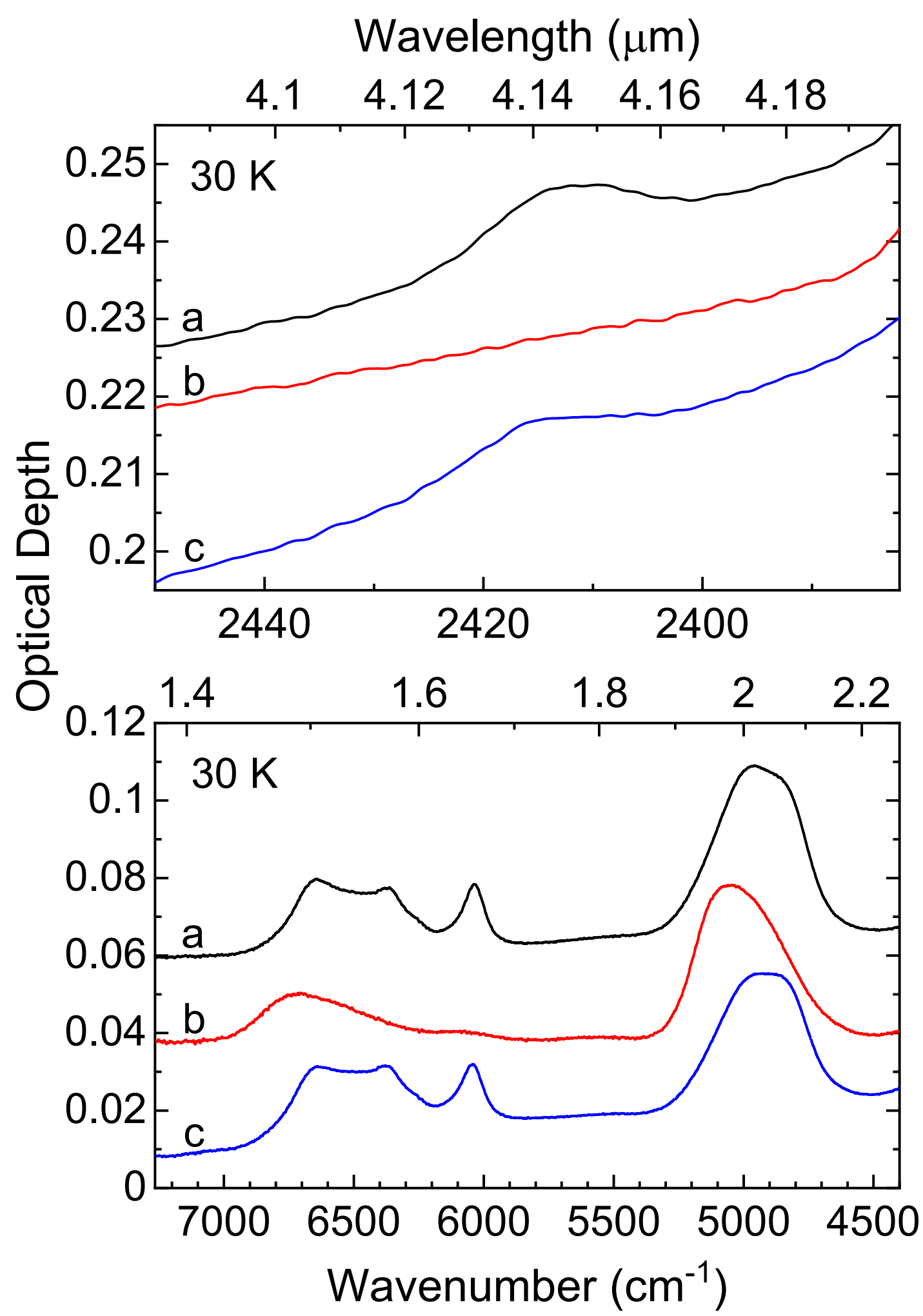

Figure 5. Infrared spectra of the HDO absorption band (top) and $H_2O$ absorption bands (bottom) of a 6.18 x $10^{18}$ $H_2O$ $cm^{-2}$ (2.17 µm) crystalline sample before (a), after irradiating with 10 keV electrons to a fluence of 1x $10^{17}$ electrons $cm^{-2}$ at 30 K (b), and after heating the irradiated sample to 160 K and recooling it to 30 K (c). The spectra have been vertically offset for clarity.

## 3. Astronomical Implications

Our results show that the 4.14 µm HDO absorption band is observable in crystalline $H_2O$-ice samples but decreases rapidly at a rate consistent with the amorphization of the sample during radiation processing at 50 K. Thus, unless the surface of interest is entirely crystalline, recent approaches using the HDO band (Clark, et al. 2019; Brown, et al. 2025; Brown, et al. 2026; Tegler, et al. 2026) to estimate the D/H ratio could lead to underestimates of the surface deuterium abundance.

So far, HDO has been primarily detected in the Saturnian (Clark, et al. 2019; Brown, et al. 2025) and Uranian systems (Brown, et al. 2026; Tegler, et al. 2026). Although these objects have surface temperatures that reach ≲100 K (e.g., (Hanel et al. 1986; Grundy et al. 1999; Filacchione et al. 2016), making their surfaces susceptible to amorphization by energetic particles (Strazzulla et al. 1992; Loeffler, et al. 2020), they still appear to have a significant crystalline component (Grundy, et al. 1999; Cruikshank et al. 2005; Emery et al. 2005; Brown, et al. 2006; Clark et al. 2012; Clark et al. 2013; Clark et al. 2014; Hedman, et al. 2024). Interestingly, Brown et al. (2025) found that the D/H ratio was lower on the trailing hemispheres of Saturn's tidally-locked moons, which we propose indicates that the fraction of crystalline $H_2O$ is slightly lower on the trailing hemispheres.

Finding more amorphous $H_2O$ on the trailing hemispheres than on the leading hemispheres is consistent with measurements of Rhea and Dione (Dalle Ore, et al. 2015; Dalle Ore, et al. 2021), and with what we know about the Saturnian radiation environment. While the leading and trailing hemispheres of Saturn's tidally-locked moons are irradiated by different magnetospheric particle energies (Howett et al. 2011; Schenk et al. 2011; Paranicas et al. 2012; Paranicas et al. 2014; Nordheim et al. 2017; Howett 2018), Loeffler et al. (2020) show that radiolytically-induced amorphization of $H_2O$-ice depends on the absorbed dose rather than the energy of individual impactors. For Mimas, the surface dose rate is about an order of magnitude higher on the trailing than on the leading hemisphere (Nordheim, et al. 2017) at the full depth probed by infrared spectroscopy, which we estimate to be 100 - 200 μm (e.g. several 1/e depths for light at 4.14 μm (Clark and Roush 1984) interacting with 50 μm $H_2O$-ice grains (Filacchione et al. 2012)). Within those depths, the 1 eV molecule$^{-1}$ dose needed to cause the 4.14 μm feature to decrease by a factor of two would be reached in only ~$1 \times 10^4$ yrs on the leading hemisphere of Mimas (Nordheim, et al. 2017). We suspect that the timescale for radiolytically-induced changes affecting the detectability of the HDO band on the other tidally-locked Saturnian satellites would be similarly short, as we have estimated for Mimas.

We can estimate the excess amorphous $H_2O$-ice required on the trailing hemispheres of the Saturnian satellites to reproduce the observed hemispherical differences in band depths reported by Brown et al. (2025) by applying the $D_{B,correction}$ factor in Figure 4. Dione and Iapetus, which show the largest hemispherical differences, require only 20% more amorphous $H_2O$-ice on their trailing hemispheres than on their leading hemispheres to match the observed band depths. This estimate is consistent with Dalle Ore et al. (2021), who found that the trailing hemisphere of Dione has ~20% more amorphous content than the leading hemisphere. Of course, if the leading hemisphere also contains amorphous ice, the estimated D/H for the entire moon may still be too low. However, if the fraction of crystalline $H_2O$-ice on the leading hemisphere can be determined from another $H_2O$ absorption feature (Hansen & McCord 2004; Cruikshank, et al. 2005; Newman et al. 2008; Dalle Ore, et al. 2015; Dalle Ore, et al. 2021), keeping in mind that sub-μm grains present on the Saturnian and possibly other icy satellites could complicate assessing the crystallinity (Clark, et al. 2012), this correction could be applied using Figure 4 before deriving the D/H ratio.

Data Availability

Upon acceptance, all data will be stored in Northern Arizona University's long-term public access data archive located at https://openknowledge.nau.edu/.

Acknowledgements

This research was supported by NASA Solar System Workings Grant # 80NSSC25K7920.